\documentclass[]{spie}

\usepackage{amsmath,amsfonts,amssymb}
\usepackage{graphicx}
\usepackage[colorlinks=true, allcolors=blue]{hyperref}

\title{Imaging performance above 150 keV of the Wide Field Monitor on board the ASTENA concept mission}

\author[a,b]{Lisa~Ferro}
\author[c,d]{Leo~Cavazzini}
\author[a]{Miguel~Moita}
\author[b]{Enrico~Virgilli}
\author[a, e]{Filippo~Frontera}
\author[b]{Lorenzo~Amati}
\author[b]{Natalia~Auricchio}
\author[b]{Riccardo~Campana}
\author[b]{Ezio~Caroli}
\author[a,b,e]{Cristiano~Guidorzi}
\author[b]{Claudio~Labanti}
\author[a,b,e]{Piero~Rosati}
\author[b]{John~B.~Stephen}
\affil[a]{Department of Physics and Earth Science, University of Ferrara, Via Giuseppe Saragat 1, Ferrara (FE), 44122, Italy}
\affil[b]{INAF/OAS Bologna, Via Piero Gobetti 101, Bologna (BO), 40129, Italy}
\affil[c]{Department of Physics, University of Trento, Via Sommarive 14, Povo (TN), 38123, Italy}
\affil[d]{Fondazione Bruno Kessler (FBK), Via Sommarive 18, Povo (TN), 38123, Italy}
\affil[e]{INFN of Ferrara, Via Giuseppe Saragat 1, Ferrara (FE), 44122, Italy}

\authorinfo{Further author information: (Send correspondence to L. Ferro)\\ Lisa Ferro: E-mail: lisa.ferro@unife.it}

\begin{document} 
\maketitle

\begin{abstract}
A new detection system for X-/Gamma-ray broad energy passband detectors for astronomy has been developed. This system is based on Silicon Drift Detectors (SDDs) coupled with  scintillator bars; the SDDs act as a direct detector of soft ($<$30 keV) X-ray photons, while hard X-/Gamma-rays are stopped by the scintillator bars and the scintillation light is collected by the SDDs. With this configuration, it is possible to build compact, position sensitive detectors with  unprecedented energy passband (2 keV -- 10/20 MeV). The X and Gamma-ray Imaging Spectrometer (XGIS) on board the THESEUS mission, selected for Phase 0 study for M7, exploits this innovative detection system.
The Wide Field Monitor - Imager and Spectrometer (WFM-IS) of the ASTENA (Advanced Surveyor of Transient Events and Nuclear Astrophysics) mission concept consists of 12 independent detection units, also based on this new technology. For the WFM-IS, a coded mask provides imaging capabilities up to 150 keV, while above this limit the instrument will act as a full sky spectrometer. However, it is possible to extend imaging capabilities above this limit by alternatively exploiting the Compton kinematics reconstruction or by using the information from the relative fluxes measured by the different cameras. In this work, we present the instrument design and results from MEGAlib simulations aimed at evaluating the effective area and the imaging performances of the WFM-IS above 150 keV.

\end{abstract}

\keywords{hard X/soft Gamma-ray astronomy, position sensitive detectors, silicon drift detectors, scintillators, point-source localization}

\section{Introduction}
\label{sec:intro}
The joint detection of a Gravitational Wave (GW) event and of a Gamma Ray Burst (GRB) has opened the era of multimessenger astrophysics\cite{Abbott17}, in which the information coming from those events will allow us to understand in an unprecedented way the most extreme phenomena and conditions in the universe, such as supernovae (SNe), GRBs, active galactic nuclei galaxies (AGNs), and even investigate the fundamental laws of physics, such as the constancy of the speed of light in vacuum. Furthermore, in the field of X and gamma-ray astrophysics, many crucial questions regarding the properties and nature of the same events are still unanswered. 

A new generation of instruments for hard X/soft gamma-ray astrophysics is necessary to reach the localization accuracy, imaging capabilities and sensitivity necessary both to investigate the high energy sky in a meaningful way and both to grant the synergy with the new ground and sky observatories and GW interferometers that will be operative in the next decades. 

\begin{figure}[t]
    \centering
    \includegraphics[scale = 0.8]{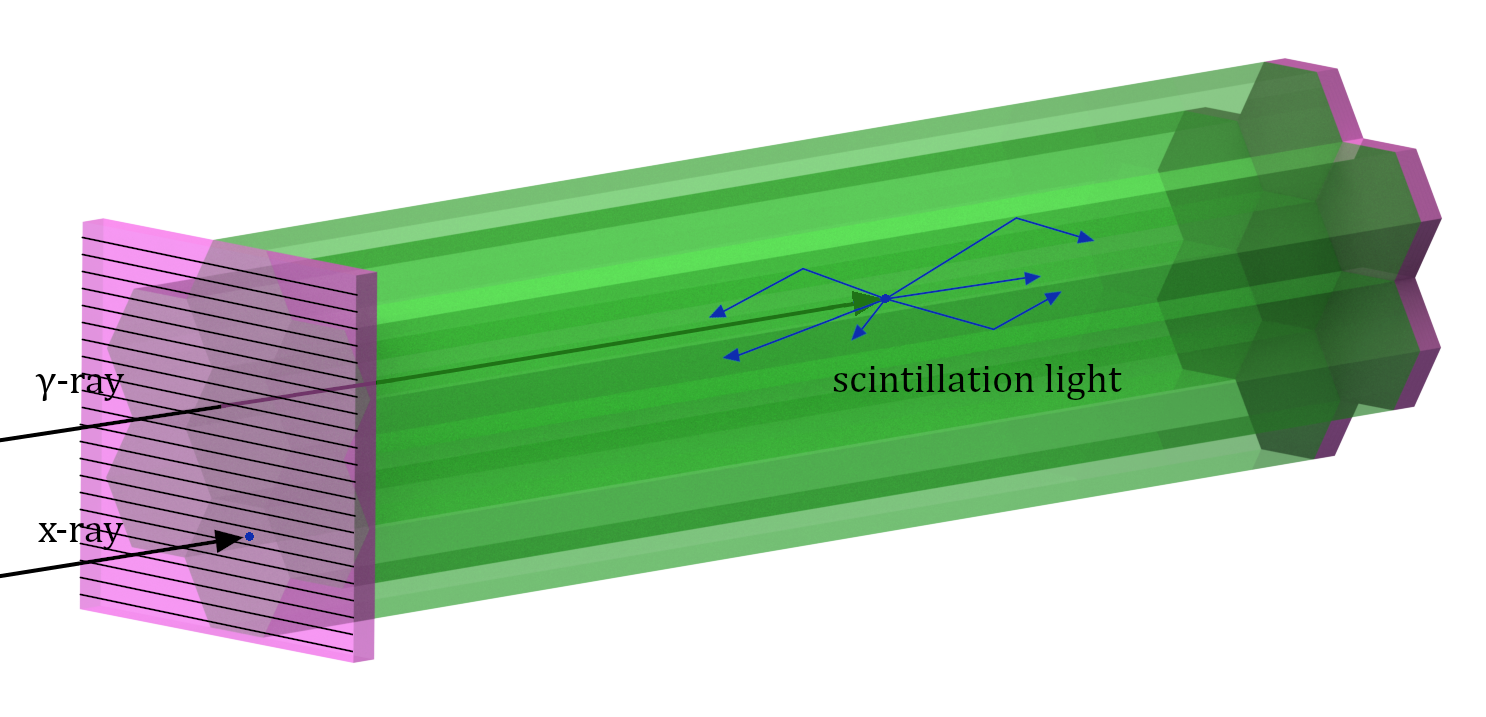}
    \caption{Representation of a small section of a full PSD unit of the WFM-IS. The scintillator bars are drawn in green, while the top and bottom SDDs in purple. Low energy X-rays are stopped by the top SDDs, while higher energy photons are stopped inside the scintillators.}
    \label{fig:siswich}
\end{figure}

With those targets in mind, a new detection system for X and Gamma ray astronomy has been developed. This detection system is based on the so-called ``siswich" (Silicon sandwich) system, which exploits the coupling between Silicon Drift Detectors (SDDs) and scintillator bars to obtain detectors working in a very broad energy band (from some keV up to tens of MeV), with 3-D position sensitivity, spectroscopic capabilities and a very low background\cite{marisaldi04}. 
In this configuration, scintillator bars are read-out on top and bottom by SDDs, as shown in Fig. \ref{fig:siswich}. Low energy ($<$ 30 keV) X-rays are stopped and detected by the SDDs on top, while higher energy gamma rays are stopped inside the scintillator bars and the SDDs, on top and bottom, detect the scintillation light. Exploiting the fact that Gamma-rays will trigger both the top and bottom SDDs, while X-ray trigger only the top SDDs, in addition to the differences in the charges pulses shapes for the two kind of events (direct detection in the SDDs and detection of scintillation light), we are able to distinguish between the two type of events.

This configuration has been proposed for the instrument XGIS (X/Gamma-ray Imaging Spectrometer) aboard the space mission THESEUS\cite{Amati2021}, selected by ESA for Phase 0 study for the M7 program. In this paper we study a possible evolution of the XGIS concept: the Wide Field Monitor - Imager and Spectrometer (WFM-IS), proposed as part of the payload for the concept mission ASTENA (Advanced Surveyor for Transient Events and Nuclear Astrophysics).

\section{The Wide Field Monitor onboard ASTENA}
\label{sec:astena}
The ASTENA concept mission was submitted to the ESA long-term program ``Voyage 2050" with two white papers\cite{Frontera2021, Guidorzi2021} describing its scope and capabilities. The ASTENA concept is based on innovative technologies and is designed not only to bring a quantum leap in terms of localization, imaging and spectroscopy in the field of X and soft gamma rays astrophysics, but also to be synergistic with the next generation of multimessenger observatories that  will be available in the decades to come.

The ASTENA payload consists in two instruments: a Narrow Field Telescope (NFT) and a Wide Field Monitor - Imager and Spectrometer (WFM-IS). The NFT will include a Laue lens, an innovative optics based on Bragg's law of diffraction, made up by thousands of crystals properly oriented to concentrate radiation in the energy band 50--700 keV to a focal point \cite{frontera2011}.

\begin{figure}[t]
    \centering
    \includegraphics[scale = 0.5]{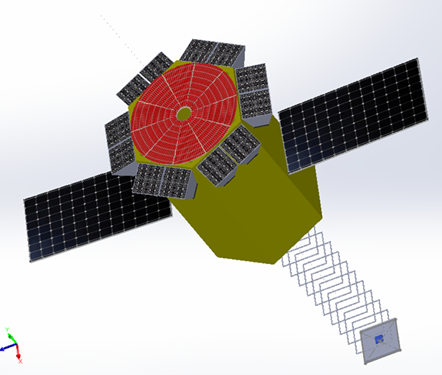}
    \caption{Schematic drawing of the ASTENA spacecraft in-flight configuration. The twelve WFM-IS Position Sensitive Modules (light grey) surround the body of the spacecraft (yellow). The optics of the Narrow Field Telescope is shown in red.}
    \label{fig:ASTENA_spacecraft}
\end{figure}

The WFM-IS, instead, will be based on the same concept of the THESEUS/XGIS, and will consist of twelve Position Sensitive Detectors (PSD) units of $43 \times 42$~cm$^2$ cross section, topped by a double scaled coded mask. The twelve PSD units will be placed around the main body of the spacecraft in groups of two, defining six different "camera pairs", oriented with an angle of 15$\mathrm{^o}$ with respect to the axis of the spacecraft (Fig. \ref{fig:ASTENA_spacecraft}).
Each PSD unit is made up by $\mathrm{4 \times 8}$ modules, each consisting of 205 hexagonal scintillator bars, with a distance between flats of 5 mm and a length that will be optimized to grant the best localization and spectroscopic performances. At the moment, we are studying Cesium Iodide (CsI(Tl)) scintillator bars, but we will also explore the possibility to use other scintillator materials such as GAGG(Ce). The scintillators are read-out on top by linear 0.4 mm thick multi-anode Silicon Drift Detectors (SDDs) and by hexagonal single-anode SDDs on bottom. 

The double scaled coded mask on top of each PSD unit will grant the instruments of imaging capabilities up to 150 keV, with a point source localization accuracy of about 1 arcmin. This is obtained by exploiting the combined pattern of a 1-D stainless steel coded mask 0.5 mm thick, for lower energy ($<$30 keV) photons, and a 2-D tungsten mask 1.0 mm thick, for higher energy (30--150 keV) photons. Above 150 keV, the coded masks become too transparent and the only way to get a rough point source localization is either by exploiting the Compton kinematics reconstruction or performing a triangulation exploiting the relative differences between the fluxes measured on the different cameras. With this work we are trying to investigate how to perform this type of analysis and to estimate the degree of localization accuracy that we can get above the limit of 150 keV.

\section{Montecarlo model}
\label{sec:model}
\begin{figure}[t]
    \centering
    \includegraphics{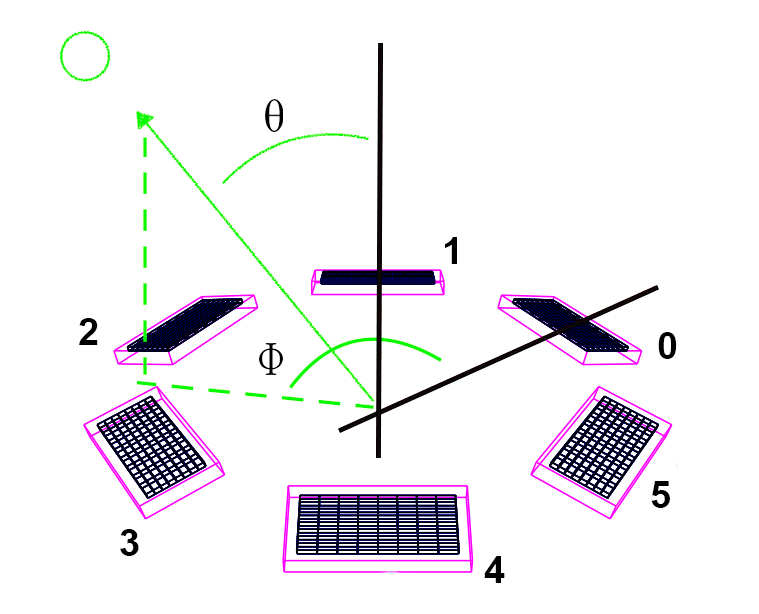}
    \caption{Monte Carlo model of the WFM-IS. It consists of a total of 12 PSD units, placed in an hexagonal shape, with 2 units on each side of the hexagon. The black grids are the instrument sensitive volumes, while the purple are vetos. The reference system and camera numbers are the ones used throughout the document.}
    \label{fig:model}
\end{figure}

A Monte Carlo model of the WFM-IS has been implemented using the MEGALib/Geomega package\cite{Cavazzini2021}. The current configuration geometry assumed for the WFM-IS is shown in Fig. \ref{fig:model};  It consists of 12 PSD units, placed in a hexagon shape, with two units on each side of the hexagon. All the units are offset by 15$^\circ$ with respect to the center axis. In this simplified geometry we did not include the coded mask and the collimator since they are almost transparent at energies $>$150 keV. Furthermore, other secondary components have not been modeled as the presence of gaps between the pixels or the wrapping material around the scintillator crystals. Such elements have a minimal impact on the results here presented and will be considered in an advanced engineering phase of the project. We included in the model the top and bottom SDDs as passive layers of Silicon. On the sides of each two units, a veto made of CsI was placed. At the moment, the simulated length of the CsI bars is 3 cm, but we will increase it to 5 cm to obtain an higher detection efficiency. The depth resolution along it is calculated using the formula on Ref.\citenum{LABANTI1991327}. In Fig. \ref{fig:model} it is also shown the reference system and camera numbers used throughout the document.

\section{Results of the simulations}
\label{sec:results}
\subsection{Effective Area and Event Multiplicity Analyses}

\begin{figure}[t]
    \centering
    \includegraphics[scale = 0.25]{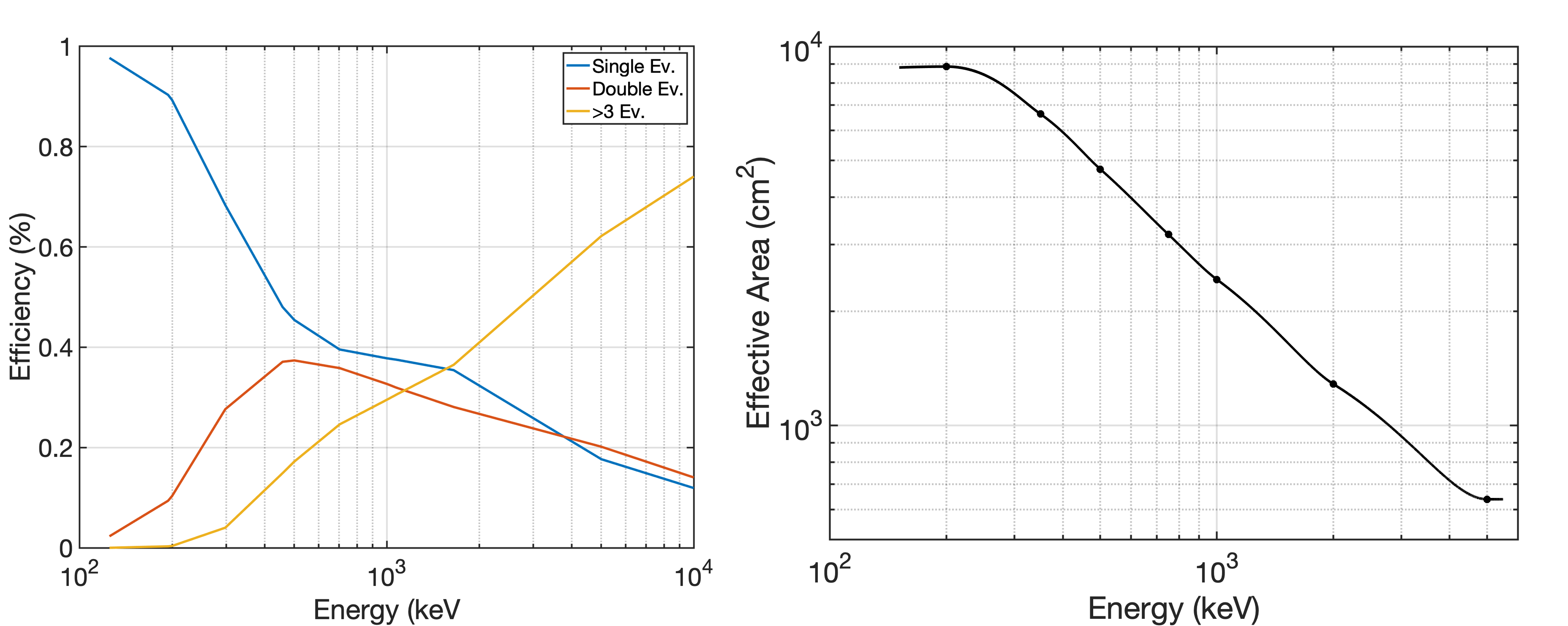}
    \caption{Left: The absolute efficiency of event with different multiplicity for the WFM-IS as a function of the energy. Right: WFM-IS effective area as a function of the energy.}
    \label{fig:MultiEffA}
\end{figure}

The detection of Compton events using the WFM-IS position sensitive detectors provides a means to localize sources. To achieve this, at least two Compton interactions are required for reconstruction. Although more than two events can enhance the localization accuracy, the benefits typically reach a point of diminishing returns, where the additional information gained from each event becomes progressively insignificant. In Fig. \ref{fig:MultiEffA} left it is shown the absolute efficiency of event multiplicity as a function of the energy for the WFM-IS. For lower energies, the single events prevail, while above $\sim$1 MeV the higher multiplicity events start to prevail.

The effective area of the WFM-IS was evaluated from simulations, both for on-axis and off-axis sources, as the ratio between the number of reconstructed Compton events and the total number of simulated events, multiplied by the surface area from which the simulated events were generated. Fig. \ref{fig:MultiEffA} right shows the effective area against the energy for an on-axis source for energies between 150~keV and 5~MeV. Between 200~keV and 5~MeV the result approximates a power law, reaching a plateau for lower and higher energies. 

\begin{figure}[t]
    \centering
    \includegraphics[scale = 0.23]{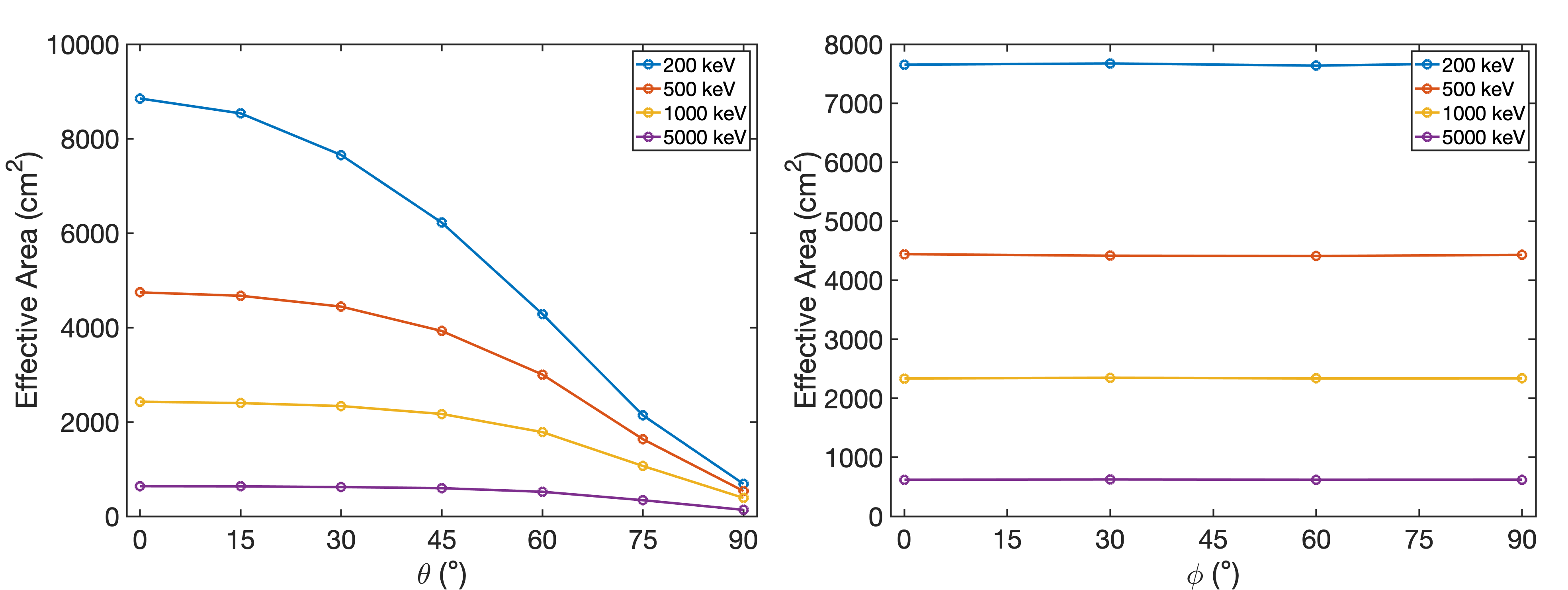}
    \caption{Left: WFM-IS effective area as function of $\theta$ for a fixed $\phi=0^{\circ}$. Right: effective area as function of $\phi$ for $\theta=30^{\circ}$.}
    \label{fig:EffA_OffAxis}
\end{figure}

The effective area for off-axis sources was calculated as a function of the zenith angle and azimuth angle, $\mathrm{\theta}$ and $\mathcal{\phi}$, shown in Fig. \ref{fig:EffA_OffAxis}. We can observe that the effective area decreases substantially with the increase of $\mathrm{\theta}$, specially for lower energies. This behavior can be exploited to localize the source. However, for higher energies, the effective area tends to be constant with $\mathrm{\theta}$, making it impossible to use this information to get the source localization. The same constant trend is found when we study the behavior of the effective area as a function of $\mathrm{\phi}$, for which the effective area remains constant independently from the value of $\mathrm{\phi}$, as observed in the right side of Fig. \ref{fig:EffA_OffAxis}, in this case for all the energies. Due to the simmetry of the cameras, the result is $\mathrm{90^{\circ}}$ symmetric, for this fact we just show the result for $\mathrm{\phi}$ between $\mathrm{0^{\circ}}$ and $\mathrm{90^{\circ}}$.

This result proves that the hexagonal positioning of the cameras is very effective to obtain a sensitivity independent on the azimuth angle; however, it is not useful for the location of the source. For that, we need to evaluate the data of each camera individually.

\subsection{Point Source Reconstruction Capabilities}
Point source reconstruction can be obtained either by studying the Compton kinematics or by exploiting the relative fluxes of the counts on the different cameras. 
Results on the Compton kinematics reconstruction are reported in the master thesis by Cavazzini (Ref. \citenum{Cavazzini2021}).

\begin{figure}[t]
    \centering
    \includegraphics[scale = 0.8]{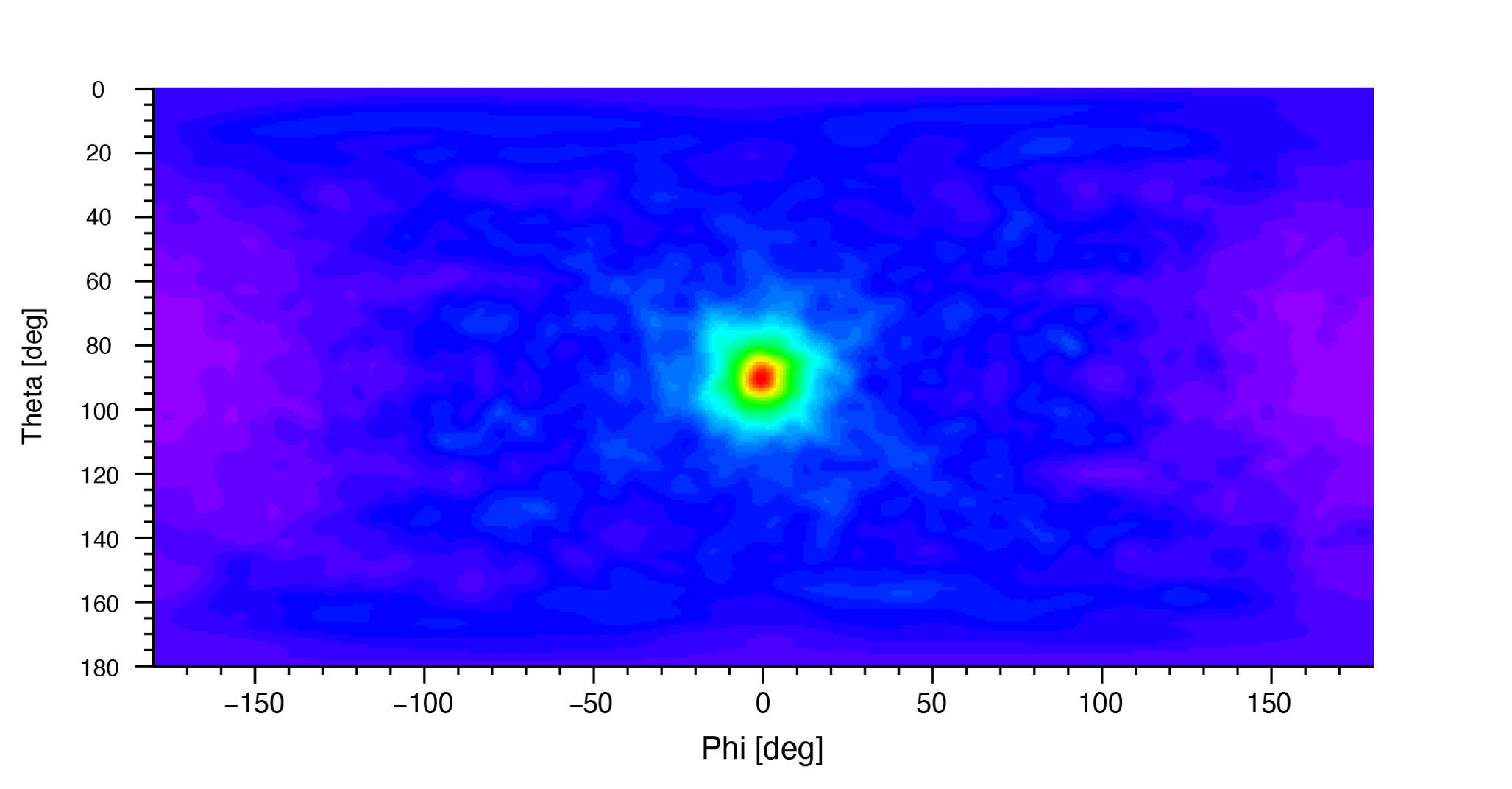}
    \caption{Reconstructed image for a on-axis 1~MeV source.}
    \label{fig:Reconst1Mev}
\end{figure}

\begin{figure}[t]
    \centering
    \includegraphics[scale = 0.2]{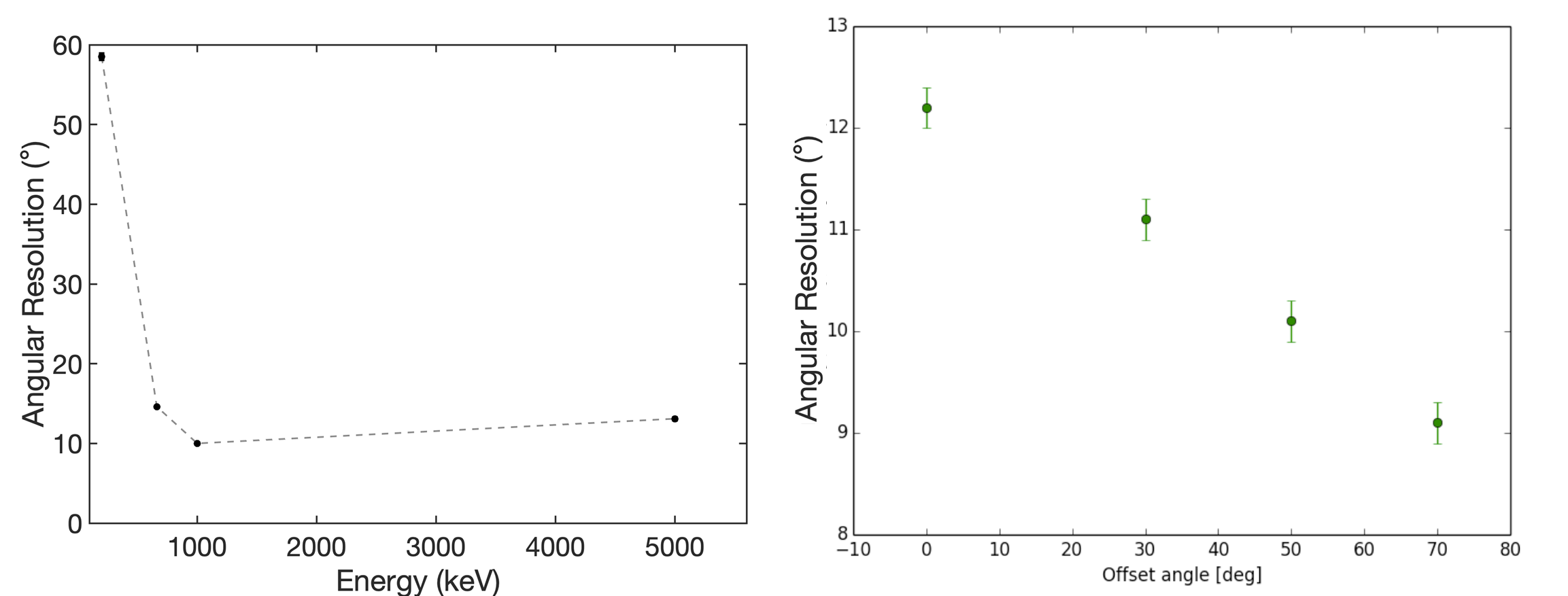}
    \caption{Left: Angular resolution as a function of energy for an on-axis source. Right: Angular resolution as a function of offset angle $\theta$ for 1~MeV source.}
    \label{fig:ARM}
\end{figure}

Scintillator PSDs are a powerful tool for Compton telescopes, allowing us to obtain measurements of the direction and energy of the photons. Their main limitation is the relatively poor energy resolution, that leads to inaccuracies in the measurement of gamma-ray energies. In Fig.~\ref{fig:Reconst1Mev} it is shown an example of an image obtained using the WFM-IS model with the MEGALib tool Mimrec for an on-axis 1 MeV source.  In Fig. \ref{fig:ARM} we also report the dependence on energy and offset angle $\mathrm{\theta}$ of the simulated Angular Resolution Measure (ARM) of the WFM-IS. The ARM is defined as the difference between the reconstructed and the true direction and it gives the accuracy on the localization of a source\cite{2006PhDTZoglauer}. The angular resolution exhibits a high value at 200 keV due to the dominance of single events, however, it shows an improvement at higher energies, since high multiplicity events become increasingly prevalent. It is also worth noticing that the angular resolution decreases when the offset angle increases. This means that the instrument reconstructs better images off axis with respect to those on axis. This is due to the camera's offset angle that allows a better reconstruction for higher angles. 

\begin{figure}[t]
    \centering
    \includegraphics[scale = 0.41]{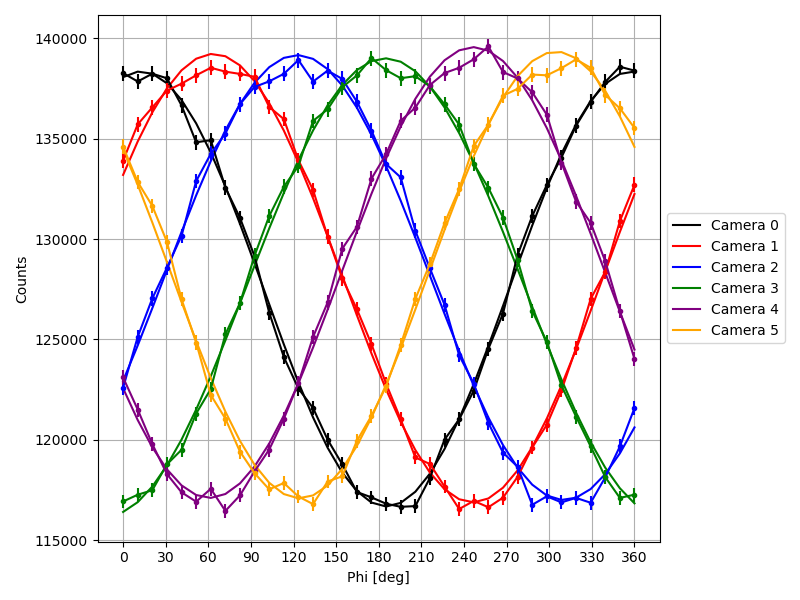}
    \includegraphics[scale = 0.41]{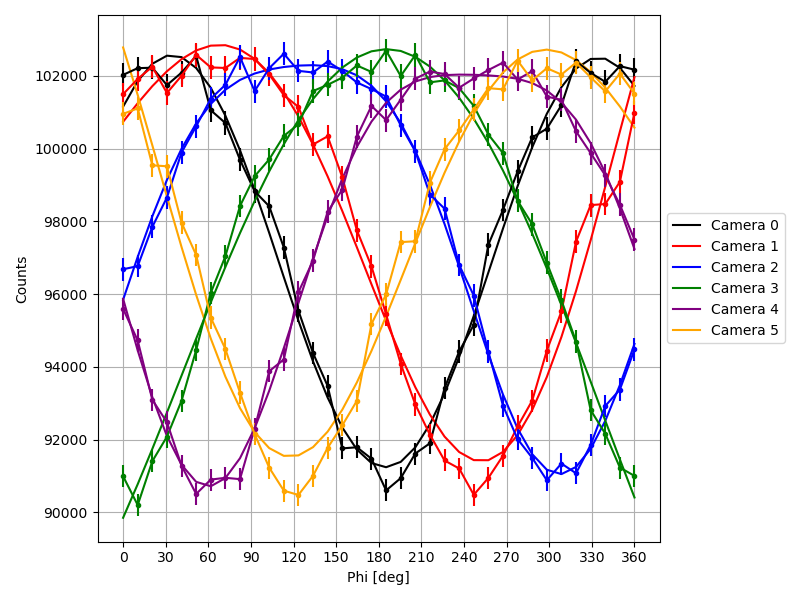}
    \caption{Counts vs azimuth angle plots for the six different camera pairs of the WFM-IS obtained with a 1 MeV beam. Left: Zenith angle ($\mathrm{\theta}$) of 15$^\circ$; right: Zenith angle ($\mathrm{\theta}$) of 40$^\circ$.}
    \label{fig:fit_curves}
\end{figure}

To obtain a very rough determination of the source's azimuth angle, we can exploit the dependency of the azimuth angle on the measured count flux on each camera pair.
We simulated a monochromatic photon wavefront with an energy of 1000 keV coming from a direction defined by the angles $\mathrm{(\theta, \phi)}$. We varied $\mathrm{\theta}$ between the values 15$^\circ$, 35$^\circ$, 40$^\circ$, 45$^\circ$, 70$^\circ$ and $\mathrm{\phi}$ in the range $\mathrm{[0^\circ,360^\circ]}$ in steps of 10$^\circ$. We fitted the counts vs $\mathrm{\phi}$ curve for each of the pair of cameras with a combination of one or more cosine functions. Figure \ref{fig:fit_curves} shows the fit on the six pairs of WFM cameras obtained for the zenith angles of 15$^\circ$ and 40$^\circ$.
We assumed the fit results as the real dependence of the counts on the cameras vs the azimuth angle of the source. We used those results to reconstruct the position of the source on the sky from the value of the counts on the six different pairs of cameras. To do so, we found the angular intervals for each pair that allow us to have a number of counts compatible with the simulated counts within five times the error on the counts. Finally, the intersection between the six angular intervals, one per camera pair, gives us an angular region corresponding to a rough localization of the source.

\begin{figure}[t]
    \centering
    \includegraphics{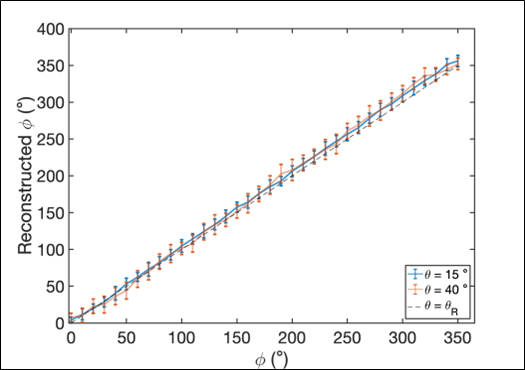}
    \caption{Simulated azimuth angle vs reconstructed azimuth angle in the case of $\theta = 15^\circ$ and $\theta = 40^\circ$. The beam was a monochromatic beam (energy = 1 MeV).}
    \label{fig:phi_rec_accuracy}
\end{figure}

Figure \ref{fig:phi_rec_accuracy} shows the results of the reconstruction of the azimuth angle (simulated azimuth angle vs reconstructed azimuth angle) for $\theta = 15^\circ$ and $\theta = 40^\circ$. The average error of the reconstruction is about 10$^\circ$. It is worth to note that the shape of the flux curves on the six camera pairs depends on the zenith angle of the source, so, with this technique, the quality of the $\theta$-localization will impact also the quality of the $\phi$-localization. 
For now, this effect was not taken in account, but we are working on understanding how the reconstruction of one angle impacts on the other. In the next future, we plan to improve this technique and test it with different values of energy, polychromatic beams and, finally, with important scientific cases. 

\acknowledgments       
 
This work has been supported with the financial contribution from the AHEAD EU Horizon 2020 project (Integrated Activities in the High Energy
Astrophysics Domain), grant agreement n. 871158.

\bibliography{biblio} 
\bibliographystyle{spiebib}

\end{document}